\documentclass[10pt]{iopart}
\usepackage{graphicx}
\usepackage{latexsym}
\usepackage{url}

\usepackage{iopams}
\def\aplett{Astrophys. Lett.\,  }
\def\apj{ApJ\,  }
\def\apjl{ApJ\,  }
\def\araa{ARA\&A  }

\def\mnras{MNRAS\,  }

\def\pasj{PASJ\,  }

\def\rmp{Rev. Mod. Phys.  }

\def\sun{\hbox{$\odot$}}
\begin{document} 
\title
{
The Fermi bubbles as a Superbubble
}
\vskip  1cm
\author     {Lorenzo  Zaninetti}
\address    {Physics Department,
 via P.Giuria 1,\\ I-10125 Turin,Italy }
\ead {zaninetti@ph.unito.it}

\begin {abstract}
In order to model the Fermi bubbles we apply the theory
of the superbubble (SB).
A thermal model and a self-gravitating model 
are reviewed.
We introduce a  third model based on the momentum conservation 
of a thin layer which propagates in a medium 
with an inverse square dependence for the density.
A  comparison have been made between  the sections of the three models 
and  the section of an observed map of  the Fermi bubbles.
An analytical law for the SB expansion as function of
the time and polar angle is deduced.
We derive a new  analytical result for the image formation of the 
Fermi bubbles in an elliptical framework.
\end{abstract}
\vspace{2pc}

\maketitle

\section{Introduction}


The term  super-shell  was observationally defined 
by \cite{heiles1979}
as holes   in the H\,I-column density   distribution  of  our Galaxy.
The dimensions of these objects  span from 100~\mbox{pc}
to 1700~\mbox{pc}
and present  elliptical shapes.
These structures  are commonly  explained through introducing
theoretical objects named  bubbles or  Superbubbles (SB);
these  are created by  mechanical energy input from stars 
(see  for example \cite{PikelNer1968}; \cite{weaver}).

The name Fermi bubbles starts to appear in the literature with
the observations of Fermi-LAT 
which  revealed  two large gamma-ray bubbles, extending
above and below the Galactic center, see 
\cite{Su2010}.
Detailed observations of the Fermi bubbles  
analyzed  
the all-sky radio region, see \cite{Jones2012}, 
the Suzaku  X-ray region, see
\cite{Kataoka2013,Tahara2015,Kataoka2015},
the  ultraviolet absorption-line spectra, see \cite{Fox2015,Bordoloi2017},
and 
the very high-energy gamma-ray emission, see \cite{Abeysekara2017}.
The existence  of the Fermi bubbles suggests 
some theoretical processes  on how they are formed.
We now outline some of them:
processes connected with the 
galactic super-massive black hole in Sagittarius A,
see \cite{Cheng2011,Yang2012,Fujita2013}.
Other studies try to explain the non thermal radiation
from Fermi bubbles in the framework of the following
physical mechanisms:
electron's  acceleration inside the bubbles,
see \cite{Thoudam2013},
hadronic  
models, 
see  \cite{Fujita2014,Cheng2015a,Sasaki2015,Keshet2017,Abeysekara2017},
and leptonic models, 
see \cite{Crocker2014,Keshet2017}.
The previous theoretical efforts allow to build a 
dynamic model for the Fermi bubbles for which the physics remains
unknown.
The layout of the paper is as follows.
In Section \ref{section_profiles} we analyze 
three  profiles in vertical density 
for the Galaxy.
In Section \ref{section_motion} we review two
existing equations of motion  for the Fermi Bubbles
and we derive a new equation of motion 
for an inverse square law in density.
In Section \ref{section_results}  we discuss
the results for the three  equations of motion
here adopted in terms of reliability of the model.
In Section \ref{section_image} we derive 
two results for the Fermi bubbles: 
an analytical  model  for the cut in intensity
in  an elliptical framework 
and a numerical map for the 
intensity of radiation  based on the  numerical section.

\section{The profiles in density}
\label{section_profiles}

This section reviews  the gas distribution in the galaxy.
A new  inverse square dependence for the gas 
in introduced.
In the following  we will use
the  spherical coordinates which are  defined  by
the radial distance $r$,
the polar angle  $\theta$,
and the azimuthal angle $\varphi$.

\subsection{Gas distribution in the galaxy}

\label{sectiongas}

The vertical density distribution
of galactic neutral atomic hydrogen (H\,I)  
is well-known; specifically, it has the following
three component  behavior as a function of
{\it z}, the  distance  from  the galactic plane in pc:
\begin{equation}
n(z)  =
n_1 e^{- z^2 /{H_1}^2}+
n_2 e^{- z^2 /{H_2}^2}+
n_3 e^{-  | z |  /{H_3}}
\,.
\label{equation:ism}
\end{equation}

We took \cite{Lockman1984,Dickey1990,Bisnovatyi1995} 
 $n_1$=0.395 ${\mathrm{particles~}}{\mathrm{cm}^{-3}}$, $H_1$=127
        \mbox{pc},
        $n_2$=0.107 $\mathrm{particles~}{\mathrm{cm}^{-3}}$, $H_2$=318
        \mbox{pc},
        $n_3$=0.064 $\mathrm{particles~}{\mathrm{cm}^{-3}}$, and  $H_3$=403
        \mbox{pc}.
This  distribution  of  galactic H\,I is valid in the range
0.4 $\leq$  $r$ $\leq$ $r_0$, where  $r_0$ = 8.5 \mbox{kpc}
and $r$  is the
distance  from  the center of the galaxy.

A recent  evaluation  for  galactic H\,I  quotes:
\begin{equation}
n_H = n_H(0) \exp{-\frac{z^2}{2 h^2}}
\quad ,
\end{equation}
with $ n_H(0)=1.11\mathrm{particles~}{\mathrm{cm}^{-3}}$,
$h=75.5\,pc$, and $z<1000\,pc$  see \cite{Zhu2017}.
A density profile of a thin
self-gravitating disk of gas which is characterized by a
Maxwellian distribution in velocity and  distribution which varies
only in the $z$-direction (ISD) has the following 
number density distribution   
\begin{equation}
n(z) = n_0 sech^2 (\frac{z}{2\,h})
\quad ,
\label{sech2}
\end{equation}
where $n_0$ is the density at $z=0$,
$h$ is a scaling parameter, 
and  $sech$ is the hyperbolic secant  
(\cite{Spitzer1942,Rohlfs1977,Bertin2000,Padmanabhan_III_2002}).

\subsection{The inverse square dependence}

The density  is  assumed to have the following
dependence on $z$
in Cartesian coordinates,
\begin{equation}
 \rho(z;z_0,\rho_0) =\rho_0\left( 1+{\frac {z}{{\it z_0}}} \right) ^{-2}
\quad .
\label{squareprofile}
\end{equation}
In the following we will adopt the following 
density profile  in spherical coordinates
\begin{equation}
 \rho(z;z_0,\rho_0) =
\cases{\rho_{{0}}&$r<r_{{0}}$\cr {\rho_{{0}} \left( 1+{\frac {r\cos \left(
\theta \right) }{z_{{0}}}} \right) ^{-2}}&$r_{{0}}<r$\cr}
\end{equation}
where the parameter $z_0$ fixes the scale and  $\rho_0$ is the
density at $z=z_0$.
Given a solid angle  $\Delta \Omega$
the mass $M_0$ swept
in the interval $[0,r_0]$
is
\begin{equation}
M_0 =
\frac{1}{3}\,\rho_{{0}}\,{r_{{0}}}^{3} \,\Delta \Omega
\quad .
\end{equation}
The total mass swept, $M(r;r_0,z_0,\theta,\rho_0,\Delta \Omega) $,
in the interval $[0,r]$
is
\begin{eqnarray}
M(r;r_0,z_0,\theta,\rho_0,\Delta \Omega)=
\nonumber \\ 
\Biggr (
\frac{1}{3}\,\rho_{{0}}{r_{{0}}}^{3}+{\frac {\rho_{{0}}{z_{{0}}}^{2}r}{
 \left( \cos \left( \theta \right)  \right) ^{2}}}-2\,{\frac {\rho_{{0
}}{z_{{0}}}^{3}\ln  \left( r\cos \left( \theta \right) +z_{{0}}
 \right) }{ \left( \cos \left( \theta \right)  \right) ^{3}}}
\nonumber \\
-{\frac {
\rho_{{0}}{z_{{0}}}^{4}}{ \left( \cos \left( \theta \right)  \right) ^
{3} \left( r\cos \left( \theta \right) +z_{{0}} \right) }}-{\frac {
\rho_{{0}}{z_{{0}}}^{2}r_{{0}}}{ \left( \cos \left( \theta \right) 
 \right) ^{2}}}+2\,{\frac {\rho_{{0}}{z_{{0}}}^{3}\ln  \left( r_{{0}}
\cos \left( \theta \right) +z_{{0}} \right) }{ \left( \cos \left( 
\theta \right)  \right) ^{3}}}
\nonumber  \\
+{\frac {\rho_{{0}}{z_{{0}}}^{4}}{
 \left( \cos \left( \theta \right)  \right) ^{3} \left( r_{{0}}\cos
 \left( \theta \right) +z_{{0}} \right) }}
\Biggl )   
\Delta \Omega
\quad .
\label{mass_square}
\end{eqnarray}
The density $\rho_0$ can be  obtained
by introducing  the number density  expressed  in particles
$\mathrm{cm}^{-3}$,
$n_0$,
the mass of  hydrogen, $m_H$,
and  a multiplicative factor $f$,
which is chosen to be  1.4, see \cite{Dalgarno1987},
\begin{equation}
\rho_0  = f  m_H n_0
\quad .
\end{equation}
An  astrophysical version of the total swept mass,
expressed in solar mass
units, $M_{\sun}$, can be obtained
introducing 
$z_{0,pc}$, $r_{0,pc}$ and $r_{0,pc}$
which are   $z_0$,      $r_0$      and $r$ expressed  in pc units.

\section{The equation of motion}
\label{section_motion}

This section reviews the equation of motion 
for  a thermal model
and for a  recursive cold model.
A new equation of motion for  a thin layer 
which propagates in a medium 
with an  inverse square dependence for the density is analyzed.

\subsection{The thermal model}

\label{secthermal}

The starting equation for the evolution
of the SB \cite{Dyson1997,mccrayapj87,Zaninetti2004}
 is  momentum conservation
 applied to a pyramidal section.
The  parameters of the thermal model are 
$N^*$,
the number of SN explosions in  $5.0 \cdot 10^7$ \mbox{yr},
$z_{\mathrm{OB}}$,
the distance of the OB associations from the galactic plane,
$E_{51}$, 
the  energy in  $10^{51}$ \mbox{erg},
$v_0$, 
the initial velocity which is fixed by the bursting phase,
$t_0$,
the initial time in $yr$  which is equal to the bursting time,
and $t$  the proper time  of the SB.
The SB evolves in a standard  three component medium,
see formula (\ref{equation:ism}).

\subsection{A recursive cold model}
\label{secrecursive}

The 3D expansion that starts at the origin 
of the coordinates;
velocity and radius are  given
by a recursive relationship,
see \cite{Zaninetti2012g}.
The parameters are the same of  
the thermal model  and  the SB evolves in a 
self-gravitating medium as given by equation (\ref{sech2}).

\subsection{The inverse square model}

\label{secinversesquare}
In the case of an inverse square  density profile
for the interstellar medium  ISM 
as given by equation (\ref{squareprofile}),
the differential equation
which models momentum conservation
is
\begin{eqnarray}
 \Biggl ( \frac{1}{3}\,\rho_{{0}}{r_{{0}}}^{3}+{\frac {\rho_{{0}}{z_{{0}}}^{2}r
 \left( t \right) }{ \left( \cos \left( \theta \right)  \right) ^{2}}}
-2\,{\frac {\rho_{{0}}{z_{{0}}}^{3}\ln  \left( r \left( t \right) \cos
 \left( \theta \right) +z_{{0}} \right) }{ \left( \cos \left( \theta
 \right)  \right) ^{3}}}
\nonumber  \\
-{\frac {\rho_{{0}}{z_{{0}}}^{4}}{ \left( \cos
 \left( \theta \right)  \right) ^{3} \left( r \left( t \right) \cos
 \left( \theta \right) +z_{{0}} \right) }}
-{\frac {\rho_{{0}}{z_{{0}}}
^{2}r_{{0}}}{ \left( \cos \left( \theta \right)  \right) ^{2}}}
\nonumber \\
+2\,{
\frac {\rho_{{0}}{z_{{0}}}^{3}\ln  \left( r_{{0}}\cos \left( \theta
 \right) +z_{{0}} \right) }{ \left( \cos \left( \theta \right) 
 \right) ^{3}}}+{\frac {\rho_{{0}}{z_{{0}}}^{4}}{ \left( \cos \left( 
\theta \right)  \right) ^{3} \left( r_{{0}}\cos \left( \theta \right) 
+z_{{0}} \right) }} 
\Biggr) 
{\frac {\rm d}{{\rm d}t}}r \left( t
k \right) 
\nonumber  \\
-\frac{1}{3}\,\rho_{{0}}{r_{{0}}}^{3}v_{{0}}=0
\quad ,
\end{eqnarray}
where the initial  conditions
are  $r=r_0$  and   $v=v_0$
when $t=t_0$.
We now briefly review that
given a function $f(r)$, the Pad\'e  approximant,
after \cite{Pade1892},
is
\begin{equation}
f(r)=\frac{a_{0}+a_{1}r+\dots+a_{p}r^{o}}{b_{0}+b_{1}%
r+\dots+b_{q}r^{q}}
\quad ,
\end{equation}
where the notation is the same of \cite{NIST2010}.
The coefficients $a_i$ and $b_i$
are found through Wynn's cross rule,
see \cite{Baker1975,Baker1996}
and our choice is $o=2$ and $q=1$.
The choice of  $o$ and $q$ is a compromise between
precision, high values for  $o$ and $q$, and
simplicity of the expressions to manage,
low values for  $o$ and $q$.
The inverse of the velocity 
is
\begin{equation}
(\frac{1}{v(r)}) = \frac{NN}{DD}
\quad ,
 \end{equation}
 where 
\begin{eqnarray}
NN=   
\left( \cos \left( \theta \right)  \right) ^{5}{r_{{0}}}^{4}r+
 \left( \cos \left( \theta \right)  \right) ^{4}{r_{{0}}}^{4}z_{{0}}+
 \left( \cos \left( \theta \right)  \right) ^{4}{r_{{0}}}^{3}rz_{{0}}
\nonumber \\
+
 \left( \cos \left( \theta \right)  \right) ^{3}{r_{{0}}}^{3}{z_{{0}}}
^{2}
-3\, \left( \cos \left( \theta \right)  \right) ^{3}{r_{{0}}}^{2}r
{z_{{0}}}^{2}+3\, \left( \cos \left( \theta \right)  \right) ^{3}r_{{0
}}{r}^{2}{z_{{0}}}^{2}
\nonumber \\
-6\, \left( \cos \left( \theta \right)  \right) 
^{2}\ln  \left( r\cos \left( \theta \right) +z_{{0}} \right) r_{{0}}r{
z_{{0}}}^{3}
\nonumber  \\
+6\, \left( \cos \left( \theta \right)  \right) ^{2}\ln 
 \left( r_{{0}}\cos \left( \theta \right) +z_{{0}} \right) r_{{0}}r{z_
{{0}}}^{3}-3\, \left( \cos \left( \theta \right)  \right) ^{2}{r_{{0}}
}^{2}{z_{{0}}}^{3}
\nonumber \\
+3\, \left( \cos \left( \theta \right)  \right) ^{2}
{r}^{2}{z_{{0}}}^{3}-6\,\cos \left( \theta \right) \ln  \left( r\cos
 \left( \theta \right) +z_{{0}} \right) r_{{0}}{z_{{0}}}^{4}
\nonumber \\
-6\,\cos
 \left( \theta \right) \ln  \left( r\cos \left( \theta \right) +z_{{0}
} \right) r{z_{{0}}}^{4}+6\,\cos \left( \theta \right) \ln  \left( r_{
{0}}\cos \left( \theta \right) +z_{{0}} \right) r_{{0}}{z_{{0}}}^{4}
\nonumber \\
+6
\,\cos \left( \theta \right) \ln  \left( r_{{0}}\cos \left( \theta
 \right) +z_{{0}} \right) r{z_{{0}}}^{4}-6\,\cos \left( \theta
 \right) r_{{0}}{z_{{0}}}^{4}+6\,\cos \left( \theta \right) r{z_{{0}}}
^{4}
\nonumber \\
-6\,\ln  \left( r\cos \left( \theta \right) +z_{{0}} \right) {z_{{0
}}}^{5}+6\,\ln  \left( r_{{0}}\cos \left( \theta \right) +z_{{0}}
 \right) {z_{{0}}}^{5}
\end{eqnarray}
and
\begin{eqnarray}
DD=  \nonumber \\
{r_{{0}}}^{3}v_{{0}} \left( \cos \left( \theta \right)  \right) ^{3}
 \left( rr_{{0}} \left( \cos \left( \theta \right)  \right) ^{2}+\cos
 \left( \theta \right) r_{{0}}z_{{0}}+\cos \left( \theta \right) rz_{{0
}}+{z_{{0}}}^{2} \right) 
\quad .
\end{eqnarray}
 
The above result allows deducing a solution $r_{2,1}$
expressed through the Pad\`e approximant
\begin{equation}
r(t)_{2,1} = \frac{AN}{AD}
\quad , 
\label{rtpadesquare}
\end{equation}
with 
\begin{eqnarray}
AN=3\, \left( \cos \left( \theta \right)  \right) ^{2}{r_{{0}}}^{3}+2\,r_
{{0}}tv_{{0}}z_{{0}}\cos \left( \theta \right) -2\,r_{{0}}{\it t_0}\,v_
{{0}}z_{{0}}\cos \left( \theta \right) 
\nonumber \\
+10\,\cos \left( \theta
 \right) {r_{{0}}}^{2}z_{{0}}+2\,tv_{{0}}{z_{{0}}}^{2}-2\,{\it t_0}\,v_
{{0}}{z_{{0}}}^{2}-2\,r_{{0}}{z_{{0}}}^{2}
\nonumber \\
-\Bigg ( \left( r_{{0}}\cos
 \left( \theta \right) +z_{{0}} \right) ^{2} \bigg ( 9\, \left( \cos
 \left( \theta \right)  \right) ^{2}{r_{{0}}}^{4}-12\,\cos \left( 
\theta \right) {r_{{0}}}^{2}tv_{{0}}z_{{0}}
\nonumber \\
+12\,\cos \left( \theta
 \right) {r_{{0}}}^{2}{\it t_0}\,v_{{0}}z_{{0}}+4\,{t}^{2}{v_{{0}}}^{2}
{z_{{0}}}^{2}-8\,t{\it t_0}\,{v_{{0}}}^{2}{z_{{0}}}^{2}+4\,{{\it t_0}}^{
2}{v_{{0}}}^{2}{z_{{0}}}^{2}
\nonumber \\
+18\,\cos \left( \theta \right) {r_{{0}}}^
{3}z_{{0}}+42\,r_{{0}}tv_{{0}}{z_{{0}}}^{2}-42\,r_{{0}}{\it t_0}\,v_{{0
}}{z_{{0}}}^{2}+9\,{r_{{0}}}^{2}{z_{{0}}}^{2} \bigg ) \Bigg )^{1/2}
\quad ,
\end{eqnarray}
and
\begin{equation}
AD= z_{{0}} \left( 4\,r_{{0}}\cos \left( \theta \right) -5\,z_{{0}}
 \right) 
\quad .
\end{equation}
A possible  set of initial values is reported 
in Table \ref{datafitsquare} in which 
the initial  value of radius and velocity 
are fixed by the bursting phase.

\begin{table}
\caption
{
Numerical values of the parameters
for  the simulation
in the case of the inverse square model.
}
\label{datafitsquare}
 \[
 \begin{array}{lc}
 \hline
 \hline
n_0 [\frac{particles}{cm^3}] & 1      \\
E_{51}            &   1               \\
N^*               &   5.87 \,10^8     \\
 r_0              &   220 \,pc         \\
 v_0              &   3500 \frac{km}{s} \\
 z_0              &   12                \\
  t               &   5.95 \,10^7 \,yr \\
t_0               &   36948  \,yr  \\
\noalign{\smallskip}
 \hline
 \hline
 \end{array}
 \]
 \end {table}
The above parameters allows to obtain an approximate 
expansion law as function of time and polar angle
\begin{equation}
r(t)_{2,1} = \frac{BN}{BD}
\quad , 
\label{rtpadesquareastro}
\end{equation}
with 
\begin{eqnarray}
BN=
31944000\, \left( \cos \left( \theta \right)  \right) ^{2}+
 18.8632\,t\cos \left( \theta \right) + 5111040\,\cos \left( 
\theta \right) 
\nonumber \\
+ 1.0289\,t- 101376
-  \Bigg ( {   (  220
\,\cos   ( \theta   ) +12   ) ^{2}   (  21083040000\,
   ( \cos   ( \theta   )    ) ^{2}
} \nonumber \\
{
- 24899.49\,t\cos
   ( \theta   ) + 3219955200\,\cos   ( \theta   ) +
 0.0073517\,{t}^{2}
}
\nonumber  \\
{
+ 4210.27\,t- 102871295   ) } \Bigg )^{1/2}
\quad ,
\end{eqnarray}
and
\begin{equation}
BD= 
10560 \,\cos \left( \theta \right) - 720
\quad .
\end{equation}

\section{Astrophysical Results}
\label{section_results}

This section  introduces a test for 
the  reliability of the model,
analyzes the observational details of the Fermi bubbles,
reviews the results for the two models of reference
and reports the results of the  inverse square model.

\subsection{The reliability of the model}

An observational
percentage reliability, $\epsilon_{\mathrm {obs}}$,
is  introduced over the whole range
of the polar   angle  $\theta$,
\begin{equation}
\epsilon_{\mathrm {obs}}  =100(1-\frac{\sum_j |r_{\mathrm {obs}}-r_{\mathrm{num}}|_j}{\sum_j
{r_{\mathrm {obs}}}_{,j}})
\quad many\,directions 
\quad,
\label{efficiencymany}
\end{equation}
where
$r_{\mathrm{num}}$ is the theoretical radius,
$r_{\mathrm{obs}}$ is the observed    radius, 
and
the  index $j$  varies  from 1 to the number of
available observations.

\subsection{The structure of the Fermi bubbles}

\begin{figure*}
\begin{center}
\includegraphics[width=7cm]{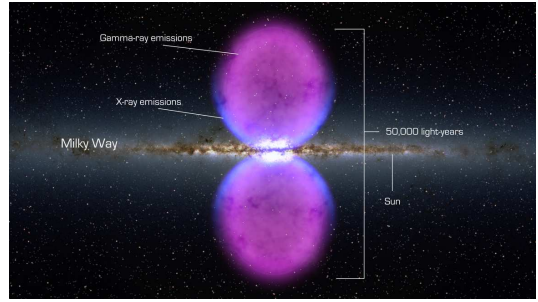}
\end {center}
\caption
{
A gamma - X image of the Fermi bubbles in 2010 as given by the NASA.  
}
\label{figbubbles}
    \end{figure*}

\begin{figure*}
\begin{center}
\includegraphics[width=7cm]{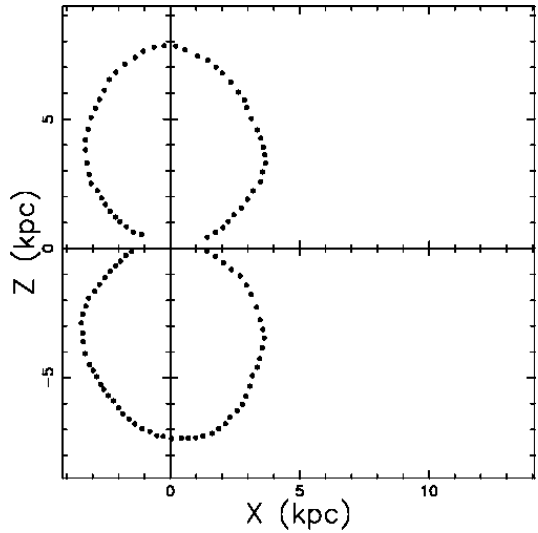}
\end {center}
\caption
{
A section of the Fermi bubbles digitalized by the author.    
}
\label{fermisb_obs}
    \end{figure*}

The exact shape of the Fermi bubbles is a 
matter of research and
as an example in  \cite{Miller2016} 
the bubbles are modeled with  
ellipsoids  centered
at  5 kpc up and below  the Galactic plane 
with semi-major axes
of 6 kpc and minor axes of 4 kpc.
In order to test our models 
we selected  the image of the Fermi bubbles 
available at
\url{https://www.nasa.gov/mission_pages/GLAST/news/new-structure.html}
which is reported in Figure~\ref{figbubbles}.
A  digitalization  of the  above advancing surface
is reported in  Figure~\ref{fermisb_obs} 
as a 2D section.
This allows to fix the observed
radii to be inserted in equation (\ref{efficiencymany}).

The actual shape of the bubbles in galactic coordinates is shown
in Figure 3 and 15 
of  \cite{Su2010}
and Figure 30 and Table 3 of  by  \cite{Ackermann2014}.

\subsection{The two models of reference}

\begin{figure*}
\begin{center}
\includegraphics[width=7cm]{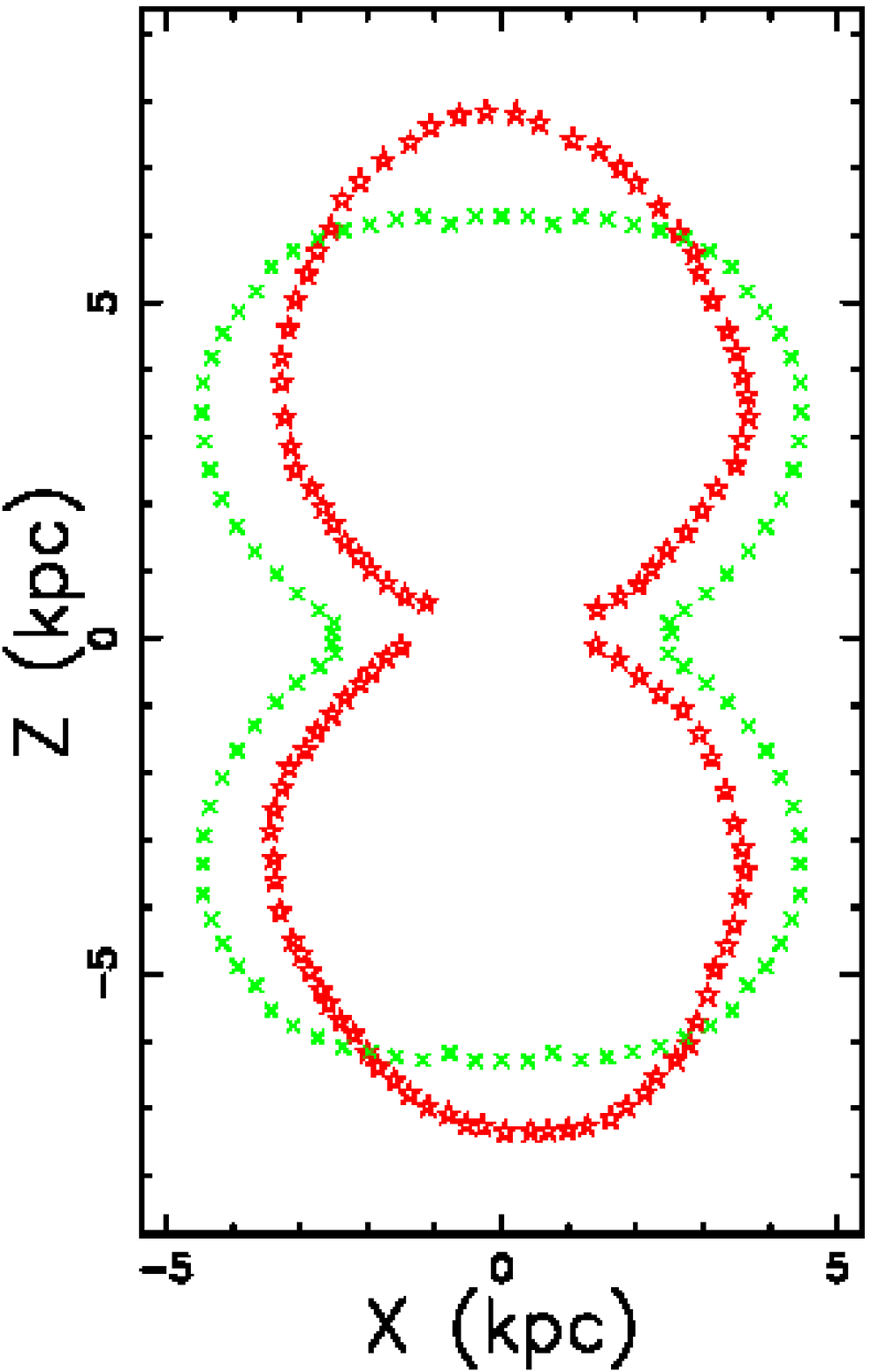}
\end {center}
\caption
{
Section of the Fermi bubbles 
in the $x-z$ plane with a thermal model
(green points)
and observed profile (red stars).
The bursting parameters
$N^*$=  113000, 
$z_{\mathrm{OB}}$=0 pc, 
$E_{51}$=1,
$t_0=0.036 \,10^7\,yr$ 
when $t=90\,10^7yr$
give
$\epsilon_{\mathrm {obs}}=73.34\%$.
}
\label{efficiency_fermib}
    \end{figure*}

The thermal model is   outlined in Section \ref{secthermal}
and Figure \ref{efficiency_fermib} reports  the numerical
solution as a  cut  in the $x-z$ plane.

\begin{figure*}
\begin{center}
\includegraphics[width=7cm]{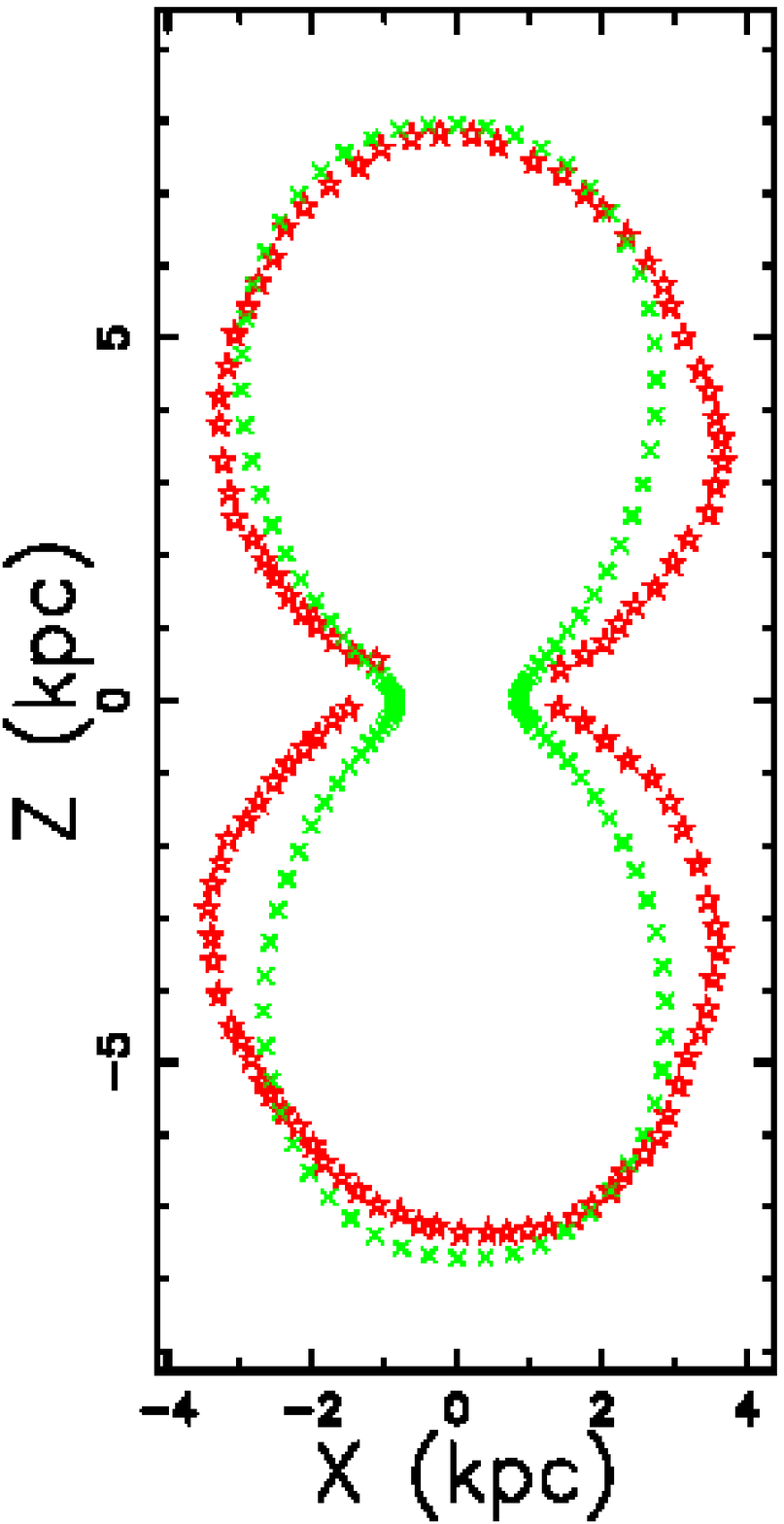}
\end {center}
\caption
{
Section of the Fermi bubbles 
in the $x-z$ plane with the cold recursive  model
(green points)
and observed profile (red stars).
The bursting parameters 
$N^*$= 79000, $z_{\mathrm{OB}}$=2 pc,
$t_0 =0.013\,10^7 $ yr  and 
$E_{51}$=1
gives 
$r_0=90.47$\ pc
and
$v_0\,=391.03$ km s$^{-1}$.
On inserting 
$h=90$\ pc,
$t=13.6\,10^7$\ yr
the reliability  
is  
$\epsilon_{\mathrm {obs}}=84.70\%$.
}
\label{section_auto}
    \end{figure*}

The cold recursive model  is   outlined in Section \ref{secrecursive}
and Figure \ref{section_auto} reports  the numerical
solution as a  cut  in the $x-z$ plane.

\subsection{The inverse square model}

\begin{figure*}
\begin{center}
\includegraphics[width=7cm]{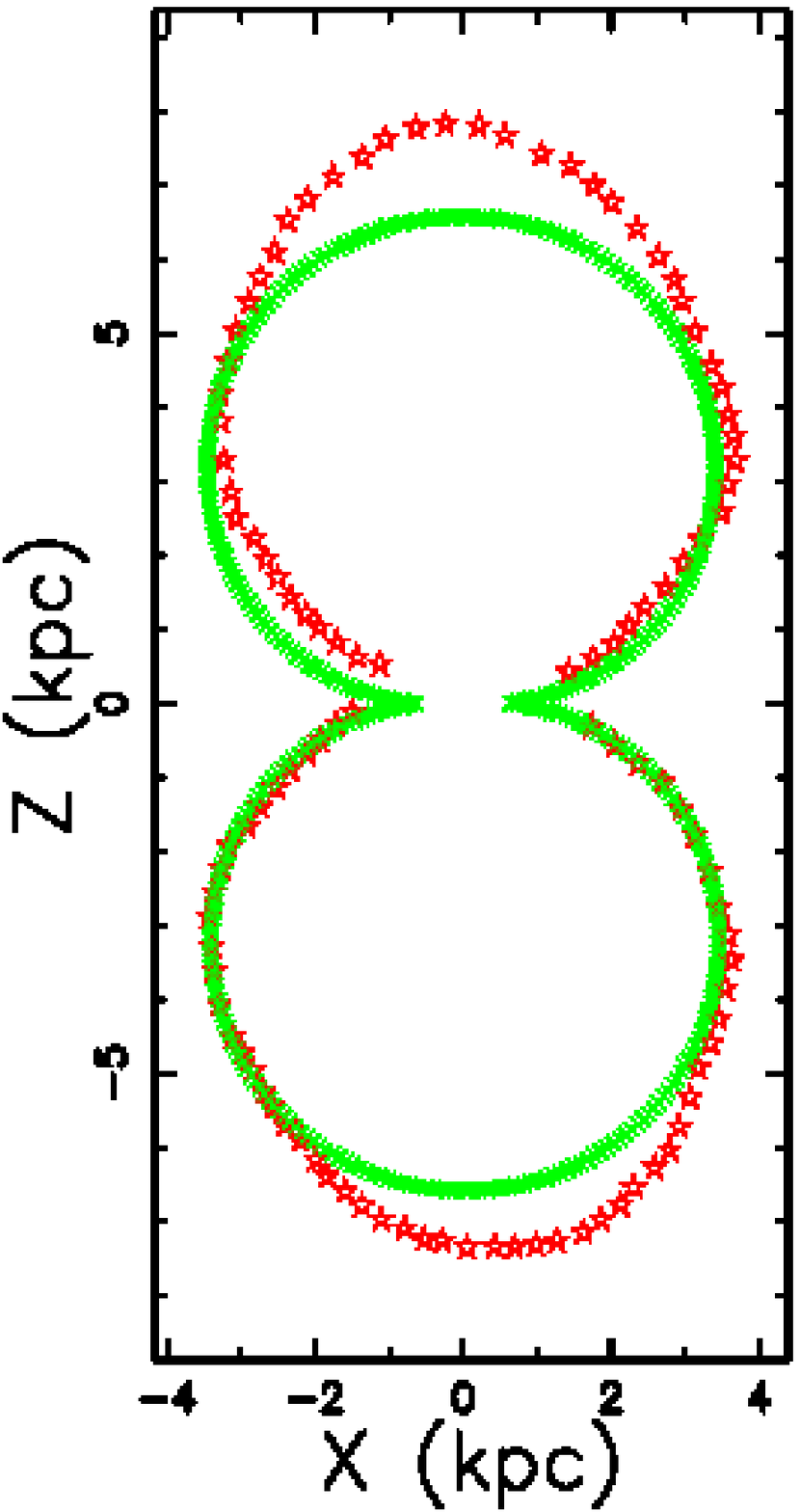}
\end {center}
\caption
{
Section of the Fermi bubbles 
in the $x-z$ plane with the inverse square  model
(green points)
and observed profile (red stars).
The  parameters  are 
reported in Table \ref{datafitsquare}
and 
the reliability 
is  
$\epsilon_{\mathrm {obs}}=90.71\%$.
}
\label{fermisb_theo_obs_square}
    \end{figure*}

The inverse square  model  is  outlined
in Section \ref{secinversesquare}
and Figure \ref{fermisb_theo_obs_square} reports  the numerical
solution as a  cut  in the $x-z$ plane.

\begin{figure*}
\begin{center}
\includegraphics[width=7cm]{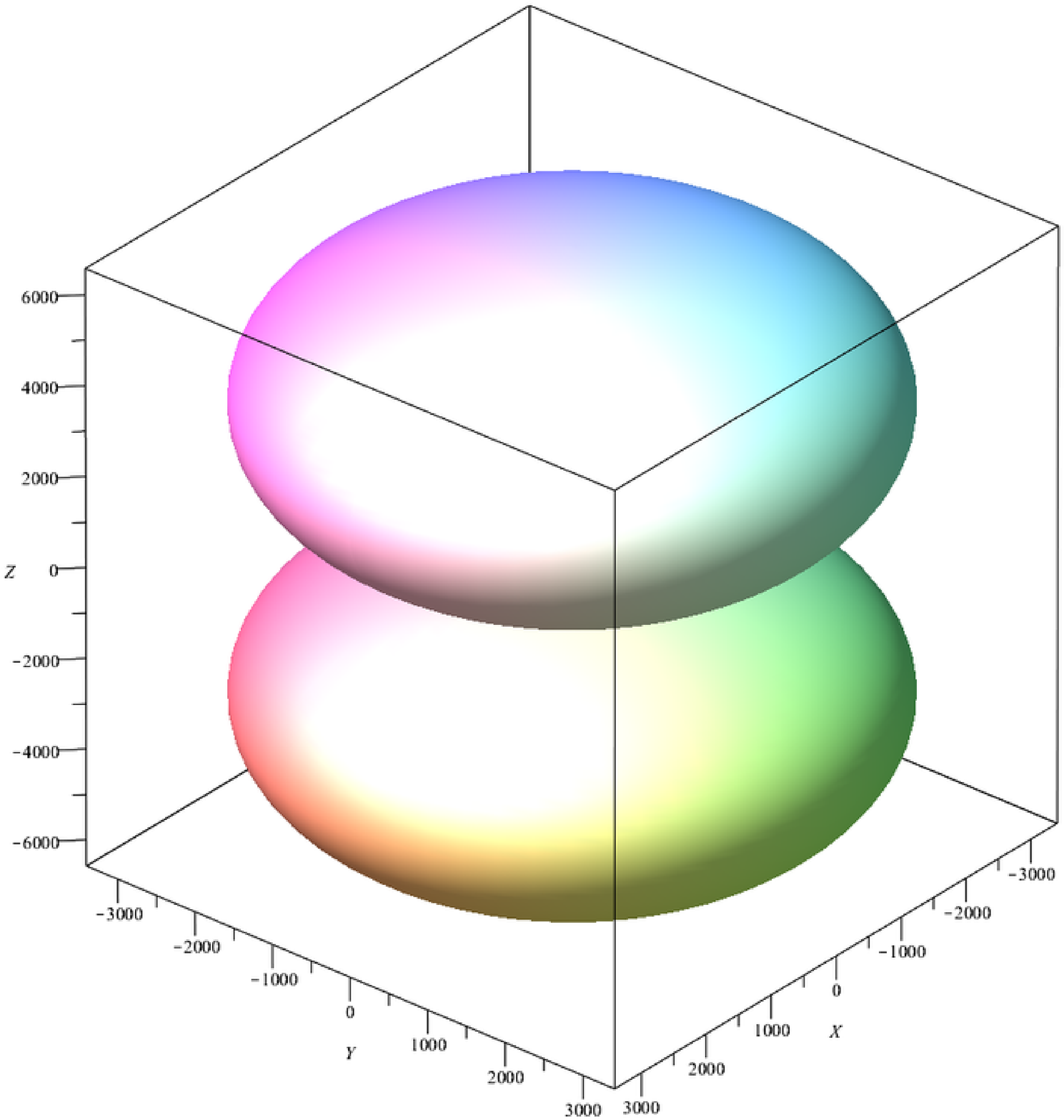}
\end {center}
\caption
{
3D surface  of the Fermi bubbles
with parameters as in Table \ref{datafitsquare}, 
inverse square  profile.
The three Euler angles are $\Theta=40$, $\Phi=60$ and
$ \Psi=60 $.
}
\label{fermisb3d}
    \end{figure*}

\begin{figure*}
\begin{center}
\includegraphics[width=7cm]{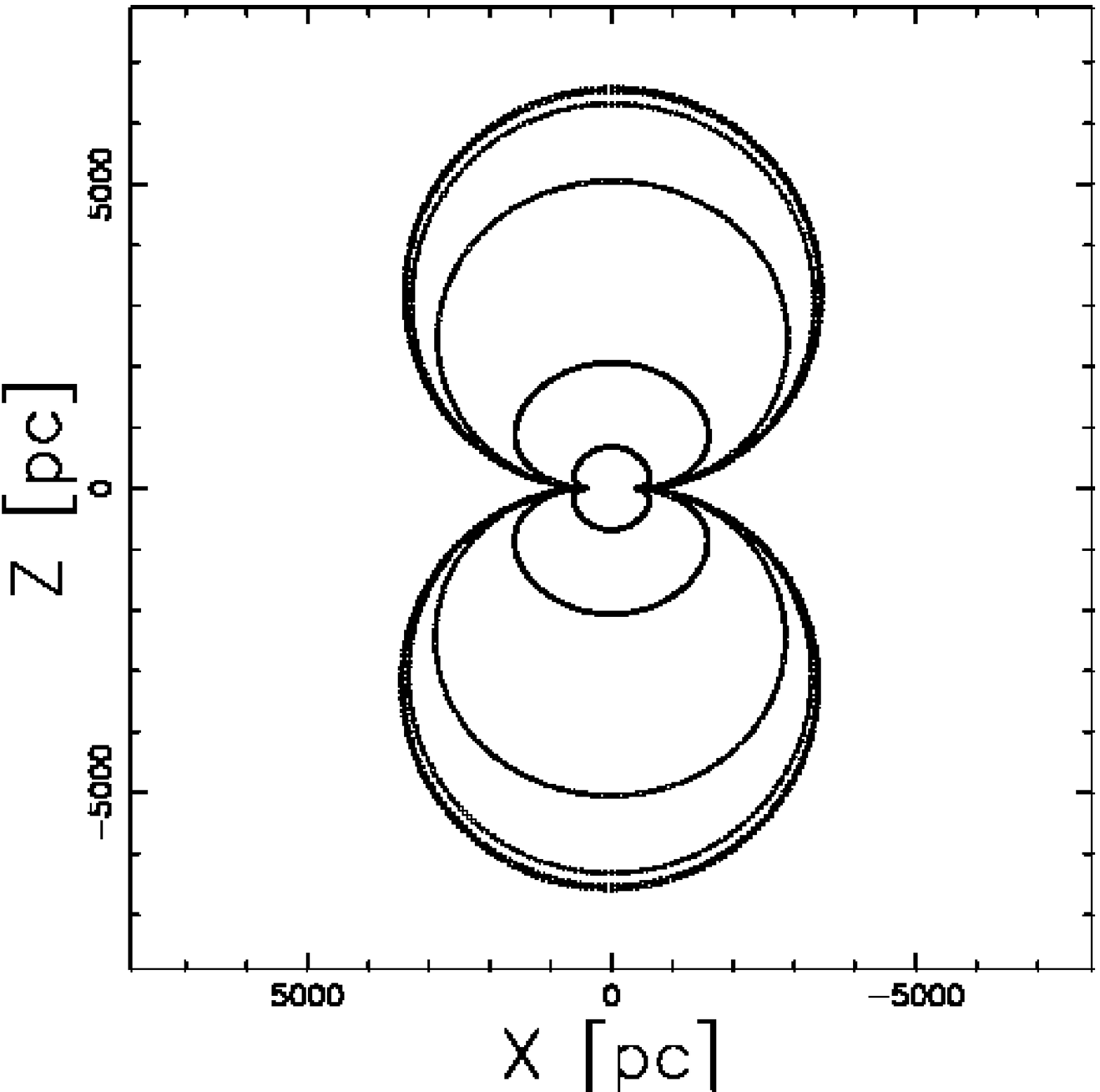}
\end {center}
\caption
{
Sections   of the Fermi bubbles
as  function of time 
with parameters as in Table \ref{datafitsquare}.
The time of each section  is    
0.0189$\,10^7\,yr$, 
0.059$\,10^7\,yr$, 
0.189$\,10^7\,yr$, 
0.6$\,10^7\,yr$, 
1.89 $\,10^7\,yr$, 
and  
6$\,10^7\,yr$.
}
\label{square_fermisb_molti}
    \end{figure*}

A rotation around  the
$z$-axis  of the above theoretical   section allows
building a 3D surface, see
Figure \ref{fermisb3d}.
The temporal evolution of the advancing surface  is reported  
in Figure \ref{square_fermisb_molti} and
a comparison should be done 
with Fig.~6  in \cite{Sofue2017}.

\section{Theory of the image}
\label{section_image}

This section reviews  
the transfer equation and 
reports a new  analytical  result for the intensity of radiation 
in an elliptical framework 
in   the non-thermal/thermal case.
A numerical model for the image formation
of   the Fermi bubbles is reported.

\subsection{The transfer equation}

The transfer equation in the presence of emission only
in the case of optically thin layer
is 
\begin{equation}
j_{\nu} \rho =K  C(s)
\quad  ,
\end{equation}
where 
$K$ is a constant, 
$j_{\nu}$ is the emission coefficient,
the index $\nu$ denotes the frequency of
emission and 
$C(s)$ is the number density of particles, 
see for example  \cite{rybicki}.
As an example  the  synchrotron emission,
as described in  sec. 4 of \cite{Schlickeiser},
is often used in order to model the radiation from  a SNR,
see for example 
\cite{Yamazaki2014,Tran2015,Katsuda2015}.
According to the above equation  the increase in intensity 
is proportional to the number density 
integrated along the line of  sight, which for constant density,
gives
\begin{equation}
I_{\nu}=K^{\prime} \times l
\quad ,
\label{equationintensity}
\end{equation}
where  $K^{\prime}$ is a constant and $l$ is the length 
along the line of sight interested in the emission;
in the case  of synchrotron emission see formula (1.175) 
in \cite{lang}. 

\subsection{Analytical non thermal model}

\begin{figure*}
\begin{center}
\includegraphics[width=7cm]{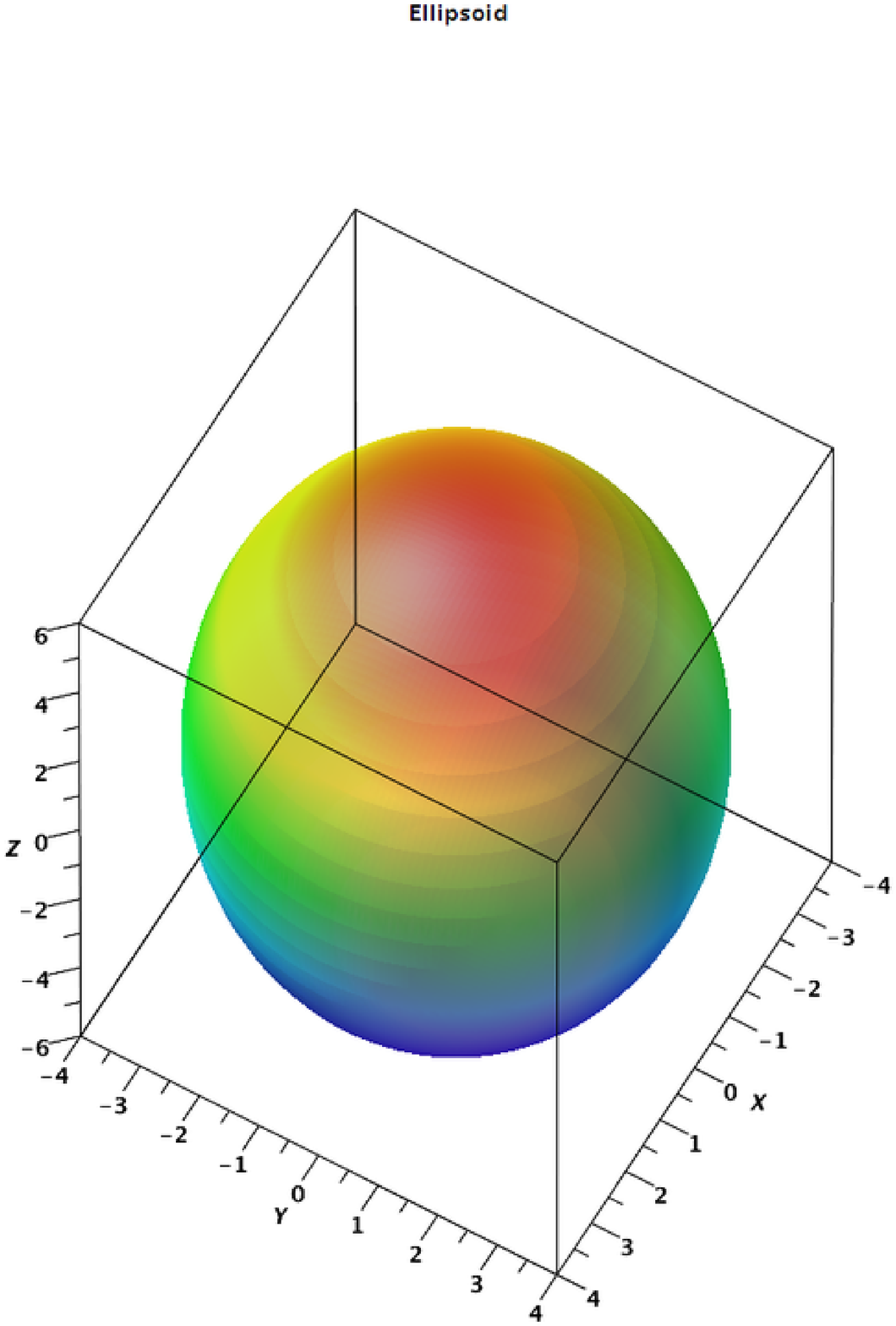}
\end {center}
\caption
{
Fermi bubbles  approximated by an ellipsoid 
when $a=6\,kpc,b=4\,kpc and d=4\,kpc$.
}
\label{ellipsoid_fermisb}
    \end{figure*}

A real ellipsoid, see \cite{Zwillinger2018},
represents a first approximation of the Fermi bubbles,
see \cite{Miller2016},
and has equation 
\begin{equation} 
\frac{z^2}{a^2} + \frac{x^2}{b^2} + \frac{y^2}{d^2}=1 
\quad ,
\label{ellipsoid}
\end{equation}
in which the polar axis  of the Galaxy  is the z-axis.
Figure \ref{ellipsoid_fermisb}  reports 
the astrophysical  application of the ellipsoid
in which due to the symmetry about the azimuthal 
angle $b=d$.

\begin{figure*}
\begin{center}
\includegraphics[width=7cm]{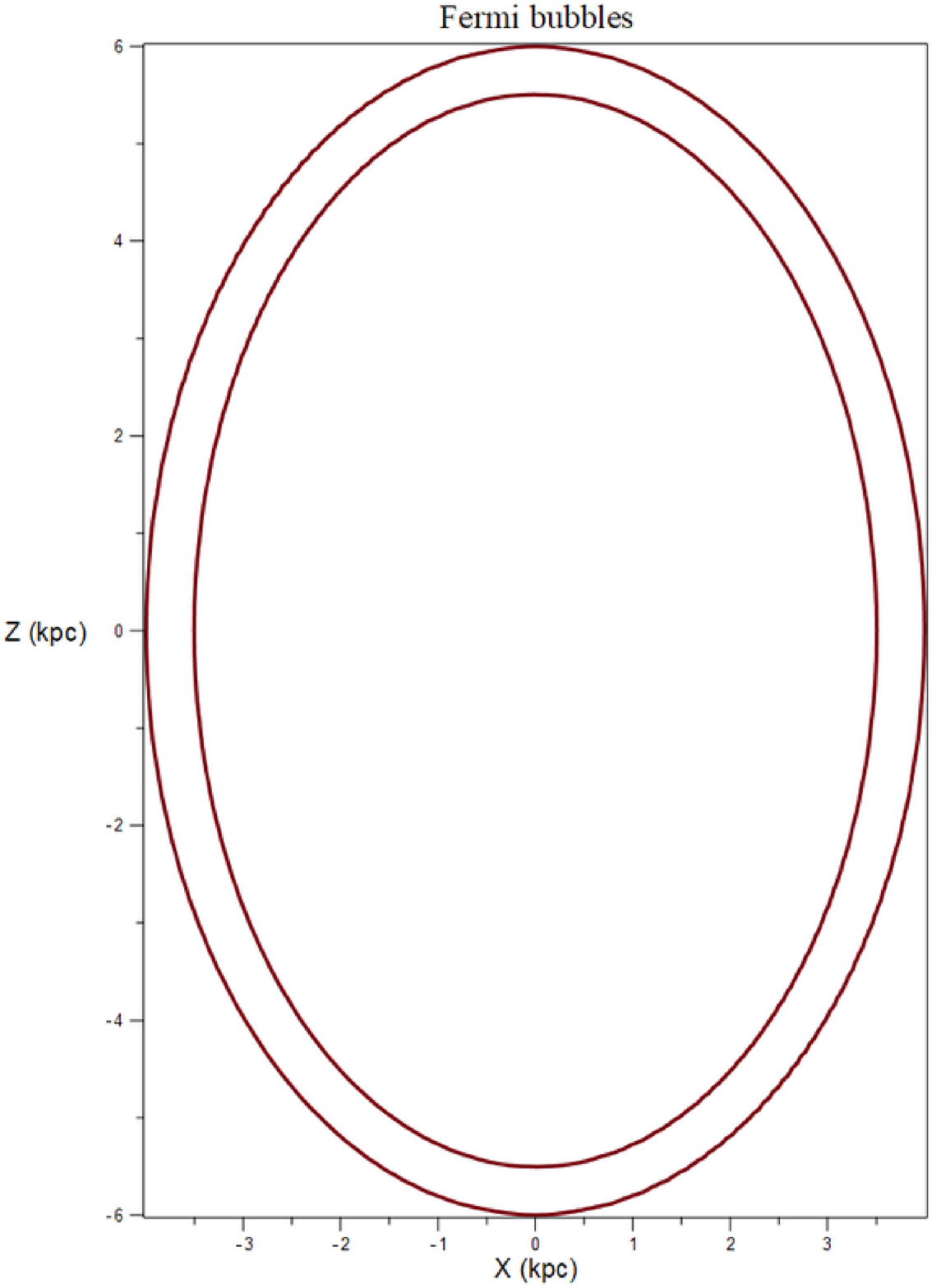}
\end {center}
\caption
{
Internal and external ellipses  
when $a=6\,kpc$,$b=4\,kpc$ and $c=\frac{a}{12}\,kpc$.
}
\label{int_ext_ellipses}
    \end{figure*}

\begin{figure*}
\begin{center}
\includegraphics[width=7cm]{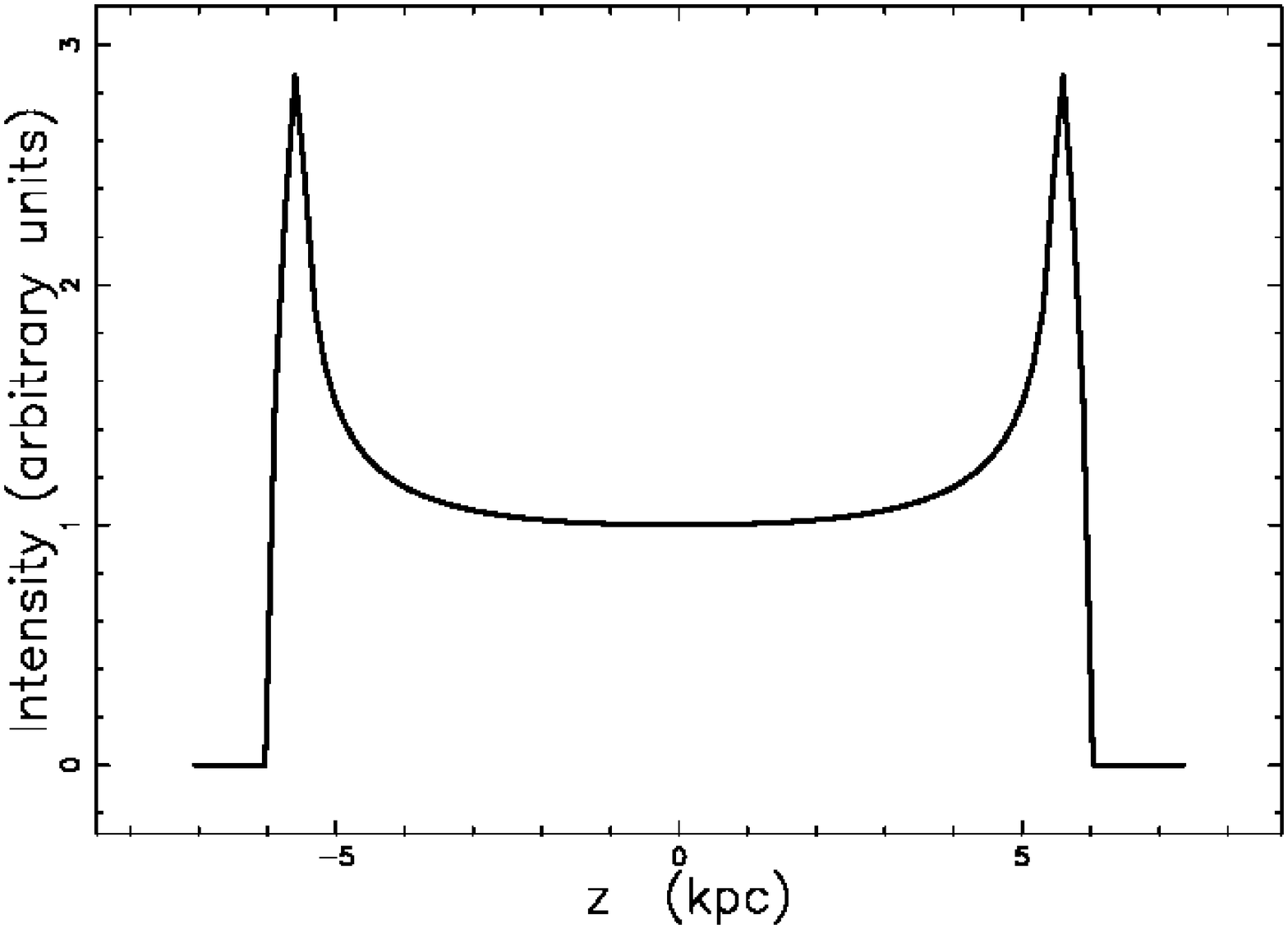}
\end {center}
\caption
{
The intensity profile along the z-axis when  
when $a=6\,kpc$,$b=4\,kpc$, $c=\frac{a}{12}\,kpc$
and $I_m$=1. 
}
\label{cut_ellipse}
    \end{figure*}

We are interested in  the section of the ellipsoid $y=0$
which is defined by the following external ellipse
\begin{equation} 
\frac{z^2}{a^2} + \frac{x^2}{b^2} =1 
\quad .
\label{ellipse}
\end{equation}
We assume 
that the emission takes place in a  thin layer comprised between
the external  ellipse
and the  internal  ellipse defined by 
\begin{equation} 
\frac{z^2}{(a-c)^2} + \frac{x^2}{(b-c)^2} =1 
\quad ,
\label{ellipseint}
\end{equation}
see Figure \ref{int_ext_ellipses}.
We therefore
assume that the number density $C$ is constant and in particular
rises from 0 at  (0,a)  to a maximum value $C_m$, 
remains constant
up to (0,a-c)  and then falls again to 0. 
The length of sight, when
the observer is situated at the infinity of the $x$-axis, is the
locus parallel to the $x$-axis which  crosses  the position $z$ in
a Cartesian $x-z$ plane and terminates at the external 
ellipse.
The locus length
is
\begin{eqnarray}
l_I  =  2\,{\frac {\sqrt {{a}^{2}-{z}^{2}}b}{a}}
\\
when \quad   (a-c) \leq z < a  \nonumber  \\
l_{II} =  
2\,{\frac {\sqrt {{a}^{2}-{z}^{2}}b}{a}}-2\,{\frac {\sqrt {{a}^{2}-2\,
ac+{c}^{2}-{z}^{2}} \left( b-c \right) }{a-c}}
\\
when \quad   
0 \leq z < (a-c)    \quad .
\nonumber  
\label{length}
\end{eqnarray}
In the case of optically thin medium, 
according to  equation (\ref{equationintensity}),
the  intensity  is split in two cases  
\begin{eqnarray}
I_I(z;a,b)  = I_m \times 2\,{\frac {\sqrt {{a}^{2}-{z}^{2}}b}{a}}
\\
when \quad   (a-c) \leq z < a  \nonumber  \\
I_{II}(z;a,,c) = I_m \times 
\Big ( 
2\,{\frac {\sqrt {{a}^{2}-{z}^{2}}b}{a}}-2\,{\frac {\sqrt {{a}^{2}-2\,
ac+{c}^{2}-{z}^{2}} \left( b-c \right) }{a-c}}
\Big )
\\
when \quad   
0 \leq z < (a-c)    \quad ,
\nonumber  
\label{intensitycut}
\end{eqnarray}
where  $I_m$  is  a constant which  allows to compare the theoretical
intensity  with the observed one.
A typical profile in intensity along the z-axis  
is reported in Figure \ref{cut_ellipse}.
The ratio, $r$, between the theoretical intensity at the maximum, $(z=a-c)$,
and at the minimum, ($z=0$), 
is given by 
\begin{equation}
\frac {I_I(z=a-c)} {I_{II}(z=0)} =r= 
{\frac {\sqrt {2\,a-c}b}{\sqrt {c}a}}
\quad .
\label{ratioteorrim}
\end{equation}
As an example the values $a=6\,kpc$,$b=4\,kpc$, $c=\frac{a}{12}\,kpc$ 
gives $r=3.19$.
The knowledge of the above ratio from the observations
allows to deduce $c$
once  $a$ and $b$ are given by the observed morphology
\begin{equation}
c =
2\,{\frac {a{b}^{2}}{{a}^{2}{r}^{2}+{b}^{2}}}
\quad .
\end{equation}
As an example in the 
 inner regions of the northeast Fermi bubble we have  $r=2$, 
see \cite{Kataoka2013}, 
which coupled with $a=6\,kpc$ and $b=4\,kpc$
gives $c=1.2\,kpc$.
The above value  is  an important astrophysical result because
we have found the 
dimension of the advancing thin layer.

\subsection{Analytical thermal model}

A thermal model for the image   is characterized by a
constant temperature in  the internal region of the advancing
section which is  approximated by an ellipse, see 
equation~(\ref{ellipse}). 
We therefore
assume that the number density $C$ is constant and in particular
rises from 0 at  (0,a)  to a maximum value $C_m$, remains constant
up to (0,-a)  and then falls again to 0.
The
length of sight, when the observer is situated at the infinity of
the $x$-axis, is the locus parallel to the $x$-axis which
crosses  the position $z$ in a Cartesian $x-z$ plane and
terminates at the external ellipse in the point (0,a).
The locus  length is
\begin{eqnarray}
l  = 2\,{\frac {\sqrt {{a}^{2}-{z}^{2}}b}{a}}  \quad  ;   -a \leq z < a
\quad . 
\end{eqnarray}
The number density $C_m$ is constant  in the ellipse 
and therefore the intensity of radiation is
\begin{eqnarray}
I(z;a,b,I_m) =I_m\times 2\,{\frac {\sqrt {{a}^{2}-{z}^{2}}b}{a}}
 \quad  ;  -a \leq z < a    \quad .
\label{ithermal}
\end{eqnarray}

\begin{figure*}
\begin{center}
\includegraphics[width=7cm]{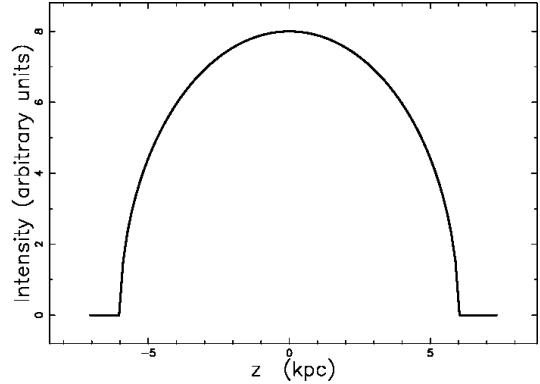}
\end {center}
\caption
{
The intensity profile along the z-axis for the thermal model 
when $a=6\,kpc$, $b=4\,kpc$  and $I_m$=1. 
}
\label{cut_ellipse_therm}
    \end{figure*}

A typical profile in intensity along the z-axis for 
the thermal model  
is reported in Figure \ref{cut_ellipse_therm}.

\subsection{Numerical model}
\begin{figure}
\includegraphics[width=6cm]{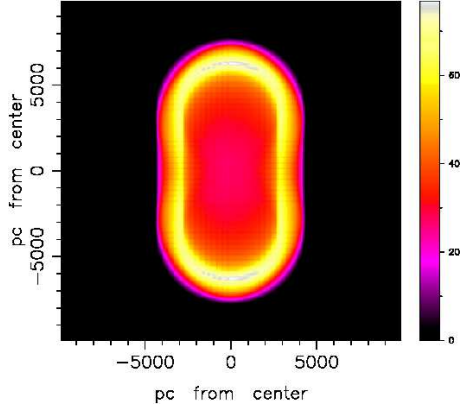}
\caption 
{
Map of the theoretical intensity  of
the Fermi bubbles 
for the 
inverse square  model
with  parameters  as 
in Table \ref{datafitsquare}.
The three Euler angles
characterizing the   orientation
  are $ \Phi $=0$^{\circ }$,
$ \Theta     $=90 $^{\circ }$
and   $ \Psi $=90  $^{\circ }$.
}
    \label{fermisb_heat}
    \end{figure}

\begin{figure}
\includegraphics[width=6cm]{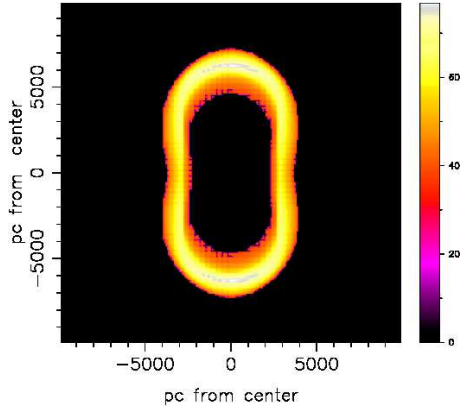}
\caption 
{
Map of the theoretical intensity  of the Fermi bubbles
as in Figure \ref{fermisb_heat}.
In this map $I_{tr}= I_{max}/2$.
}
    \label{fermisb_heat_hole}
    \end{figure}

The source of luminosity is assumed here to be
the flux of kinetic energy, $L_m$,
\begin{equation}
L_m = \frac{1}{2}\rho A  V^3
\quad,
\label{fluxkineticenergy}
\end{equation}
where $A$ is the considered area, $V$ the velocity 
and $\rho$ the density, 
see formula (A28)
in \cite{deyoung}.
In our  case $A=R^2 \Delta \Omega$,
where $\Delta \Omega$ is the considered solid angle along 
the chosen direction.
The   observed luminosity along a given direction 
can  be expressed as
\begin{equation}
L  = \epsilon  L_{m}
\label{luminosity}
\quad  ,
\end{equation}
where  $\epsilon$  is  a constant  of conversion
from  the mechanical luminosity   to  the
observed luminosity.
A  numerical algorithm which allows us to
build  a complex  image
 is 
outlined  in Section 4.1 of \cite{Zaninetti2013c}
and 
the orientation  of the object is characterized by
the
Euler angles $(\Phi, \Theta, \Psi)$
The threshold intensity can be
parametrized  to  $I_{max}$,
the maximum  value  of intensity
characterizing the map.
The image of the Fermi bubbles
is shown in Figure \ref{fermisb_heat}
and 
the introduction   of a threshold intensity
is visualized in Figure \ref{fermisb_heat_hole}.

\section{Conclusions}

{\bf Law of motion} 
We have compared two existing  models for the 
temporal evolution of the Fermi bubbles, 
a thermal model, see Section \ref{secthermal},
and an autogravitating model, see Section 
\ref{secrecursive},
with a new model which conserves the momentum 
in presence of an inverse square law  for the density of the ISM.
The best result is obtained by the inverse square 
model which produces a reliability 
of  $\epsilon_{\mathrm {obs}}=90.71\%$ for the expanding radius
in respect to a digitalized section of  the Fermi bubbles.
A semi-analytical  law of motion as  function 
of polar angle  and  time is derived 
for  the inverse square model, 
see equation (\ref{rtpadesquareastro}).

{\bf Formation of the image}
An analytical cut for the intensity of radiation 
along the z-axis  is derived in the framework of
advancing surface  characterized  by an internal 
and an external ellipses.
The analytical cut in theoretical intensity 
presents a characteristic "U" shape which has  
a maximum in the  external ring and a 
minimum at the center, see equation~(\ref{intensitycut}).
The presence of a hole  in the intensity of radiation
in the central region of the elliptical Fermi bubbles
is also confirmed by a numerical algorithm for the 
image  formation, see Figure \ref{fermisb_heat_hole}.
The theoretical  prediction of a hole in the intensity 
map explains the decrease  in intensity 
for  the 0.3 kev plasma by ~= $50\%$
 toward
the central region of the northeast Fermi bubble, see 
\cite{Kataoka2013}.
The intensity of radiation for the
thermal model conversely  presents a  maximum of the intensity at the
center of the elliptical Fermi bubble,
see equation (\ref{ithermal}) 
and this theoretical prediction does not agree 
with the above observations.

\section*{Acknowledgments}
Credit for 
Figure \ref{figbubbles} is given
to NASA.
  
\leftline{\bf References}

\providecommand{\newblock}{}

\providecommand{\newblock}{}
\label{lastpage}
\end{document}